\documentclass[]{raa}            % referee version: for submission
\usepackage{graphicx,times}
\usepackage{natbib}

\begin{document}

   \title{RXTE/ASM and Swift/BAT Observations of Spectral Transitions in Bright X-ray Binaries in 2005--2010
% $^*$
%\footnotetext{\small $*$ Supported by the National Natural Science Foundation of China.}
}

 \volnopage{ {\bf 2010} Vol.\ {\bf x} No. {\bf XX}, 000--000}
   \setcounter{page}{1}

   \author{Jing Tang
      \inst{}
   \and Wenfei Yu
      \inst{}
   \and Zhen Yan
      \inst{}
   }

   \institute{Key Laboratory for Research in Galaxies and Cosmology, Shanghai Astronomical Observatory, 80 Nandan Road,  Shanghai 200030, China; E-mail: wenfei@shao.ac.cn \\
%        \and
%             Full institute address for the second author
%        \and
%             Full institute address for the third author
\vs \no
   {\small Received [year] [month] [day]; accepted [2010] [Nov] [25] }
}

\abstract{ We have studied X-ray spectral state transitions that can be seen in the long-term monitoring light curves of bright X-ray binaries from the All-Sky Monitor (ASM) on board the Rossi X-ray Timing Explorer (RXTE) and the Burst Alert Telescope (BAT) on board Swift during a period of five years from 2005 to 2010. We have applied a program to automatically identify the hard-to-soft (H-S) spectral state transitions in the bright X-ray binaries monitored by the ASM and the BAT. In total we identified 128 hard-to-soft transitions, of which 59 occurred after 2008. We also determined the transition fluxes and the peak fluxes of the following soft states, updated the measurements of the luminosity corresponding to the H-S transition and the peak luminosity of the following soft state in about 30 bright persistent and transient black hole and neutron star binaries following Yu \& Yan (2009), and found the luminosity correlation and the luminosity range of spectral transitions in data between 2008-2010 are about the same as those derived from the data before 2008. This further strengthen the idea that the luminosity at which the H-S spectral transition occurs in the Galactic X-ray binaries is determined by non-stationary accretion parameters such as the rate-of-change of the mass accretion rate rather than the mass accretion rate itself. The correlation is also found to hold in data of individual sources 4U 1608-52 and 4U 1636-53. 
\keywords{X-rays: binaries: state
transition } }

   \authorrunning{J. Tang et al. }            %author_head in even pages
   \titlerunning{Spectral transitions in Galactic X-ray binaries }  % title_head in odd pages
   \maketitle

%% The author head (on even pages) and the title head (on odd pages) will be
%% automatically extracted from \author{} and \title{}. Whenever the title is too long,
%% you will be asked to supply a shorter one by inserting either \authorrunning{} or
%% \titlerunning{} before \maketitle. Anyway, you can specify your own heads.
%%
%%
%% Note: In the following text body of your manuscript, please note several differences from
%%       other major journals:
%% (1) \subsection{Please Capitalize the First Letter of Each Notional Word in Subsection Title}
%% (2) Please Capitalize the First Letter of Each Notional Word in all tables' captions

%
%________________________________________________ sections below
%
\section{Introduction}           %% first-level sections will be auto-capitalized
\label{sect:intro}

Galactic black hole X-ray binaries exhibit two main X-ray
spectral states -- the hard state and the soft state (see the review Remillard \& McClintock 2006). In the soft state, the X-ray energy spectrum is dominated by a thermal component with a weak
power law tail. In the hard state, the energy is characterized by a
power law with a break or an exponential cutoff at high energy.
These two spectral states are shared by the atoll sources in the neutron
star X-ray binaries (van der Klis 1994), roughly corresponding to the banana state and
the island state, respectively (Hasinger \& van der Klis 1989), in terms of the X-ray
spectral and timing properties. The transition between the hard state and the
soft state is called as \textquotedblleft state transition \textquotedblright. In this paper, we
focus on the hard-to-soft (hereafter H-S) transition.

H-S transition can usually be seen during the rising phase of a
bright outburst of a black hole or neutron star transient, but there are exceptions. Yu \& Dolence (2007) discovered one H-S transition during a luminosity decline in Aquila X-1. A number of transient sources have been seen to stay in the hard state throughout their outbursts, e.g., XTE J1550-564 showed spectral and timing features typical of an hard state during its short outburst in January 2002 (Belloni et al. 2002). A complete H-S transition usually takes on time scales from a few days to a few weeks, varied source by source.

Up to now, the origin of the state transition has not been completely understood. The widely accepted understanding is based on stationary accretion (e.g., Esin et al. 1996). In this framework, mass accretion rate determines the state transition. However, the study of the hysteresis effect of spectral state transitions
(Miyamoto et al. 1995; Nowak 1995; Maccarone \& Coppi 2003) and the large span of H-S transition luminosities,
which could vary by one order of magnitude in a single source (Yu \& Dolence 2007), suggest
that an additional parameter is needed to interpret state evolution. Yu \& Yan (2009) showed that the rate-of-change of the mass accretion rate rather than the mass accretion rate itself drives the
state transition in most bright Galactic X-ray binaries primarily, suggesting that we need to rely on non-stationary accretion to interpret state transitions.

It has been found that the luminosity corresponding to the start of the H-S transition positively correlates with the outburst peak luminosity not only in individual transient sources Aql X-1, XTE J1550-564, and GX 339-4, and individual persistent, transient-like neutron star low-mass X-ray binary 4U 1705-44 (Yu et al. 2004, 2007; Yu \& Dolence 2007), but also in transient and persistent sources as a whole (Yu
\& Yan 2009). This correlation holds for bright outbursts as well as low-luminosity flares (Yu \& Dolence 2007; Yu \& Yan 2009), and in view of no saturation toward high luminosities, brighter hard states than the ones currently known is expected to be observed in transient sources during brighter outbursts. On the other hand, one can predict the
outburst peak luminosity during the rising phase of an outburst when
the H-S transition occurs (Yu et al. 2004, 2007; Yu \& Dolence 2007), and even earlier, during the early outburst rise using measurements of the rate-of-change of the X-ray flux (Yu \& Yan 2009).
It is therefore very important to keep tracking spectral transitions in the bright X-ray binaries --- understanding luminosity regimes of the hard state and the soft state and predicting or alerting further spectral and flux evolution.

We have systematically studied state transitions of all bright X-ray binaries, seen with X-ray monitoring observations with the All-Sky Monitor (ASM) on board the Rossi X-Ray Timing Explorer (RXTE) and
the Burst Alert Telescope (BAT) on board the Swift in the 2-12 keV and the 15-50 keV energy ranges, respectively, in a period of five years following the previous study Yu \& Yan (2009). Specifically, we applied an automatic program to search for H-S transitions in all bright X-ray binaries according to the hardness ratios between the BAT flux and the ASM flux and to determine the transition flux and the peak flux of the following soft state. We obtained the updated correlation between the luminosity corresponding to the H-S transition and the peak luminosity of the following soft state in both transient and persistent sources. In total we found 59 new H-S transitions in 2008-2010, in addition to 69 H-S transitions identified in 2005-2008.

% Authors can give a citation as `Michel et al. 1992'.
% You may also use \cite, \citep and \citet for citation, and use Table~1
% or Figure~1 and so forth. Using \ref and \label for cross-references of
% Tables/Figures is a good way in adjusting/adding/removing text, tables or
% figures.

\section{Observations and initial data analysis}
\label{sect:Obs}

Owing to the X-ray sky monitoring products from observations of the ASM (2-12 keV) and the BAT
in the energy ranges 2-12 keV and 15-50 keV, we are able to obtain daily X-ray light curves of bright
X-ray binaries. Since the energy bands of the ASM and the BAT cover
primarily for the thermal emission of the accretion disk or neutron star surface or boundary and the power-law spectral components respectively, the hardness ratios between the two instruments provide a comparison between these two spectral components, which is suitable for determining spectral states and searching for state transitions. Previous study by Yu \& Yan (2009) demonstrated the two monitors can trace spectral transitions in bright X-ray binaries very well.

We took data from 2005 February 12 (MJD 53413) to 2010 April 18 (MJD
55304) when this study was finished. Following the method described in Yu \& Yan (2009), we used
good BAT data with both data flag and dither flag being 0. The X-ray
flux was converted into units of crab --- 1 crab=75 count\ s$^{-1}$
for the ASM and 1 crab =0.23 count\ s$^{-1}$\ cm$^{-2}$ for the BAT.
To increase detection sensitivity, we calculated 2 day-averaged
results.

We studied the ASM light curve, the BAT light curve and the hardness
ratio for each source monitored by the RXTE/ASM and the Swift/BAT.
On the basis of previous results presented in Yu \& Yan (2009), we
took 1.0 and 0.2 as the hardness ratio thresholds for the hard state
and the soft state, respectively. That is, when the hardness ratios
were above 1.0, the sources were taken as in hard states and when
the hardness ratios were in the range below 0.2, the sources were
taken as in soft states. These thresholds are strict, since the observed hardness ratio for the hard state is
in the range above 0.6 and that for the soft state is below 0.35, based on the distribution of 2-day averaged hardness ratios of all the bright black hole and neutron star X-ray binaries (Yu \& Yan 2009). The later thresholds were used to estimate the significance of spectral transitions we determined in the end.

When the H-S transition occurs, the hardness ratio shifts from above
the hard state threshold to below the soft state threshold. The BAT
peak flux of the hard states around the H-S transition (within 6
days) was chosen as the transition flux corresponding to the start
of each H-S transition, since the brightest hard state during the
rising phase of transient outburst corresponds to when the transition
occurs (Yu et al. 2003, 2004). The ASM peak flux of the soft states
immediately after the H-S transition was chosen as the peak flux of
the following soft state. The significance of the peak flux we
identified for either state was required to be above $3\sigma$.
Meanwhile, we excluded those isolated ASM and BAT peaks, shown as either
outliers or only detected in a single time bin to avoid false
identifications. The detailed criteria are presented below.

   \begin{figure}[h!!!]
   \centering
   \includegraphics[width=8.0cm, angle=90]{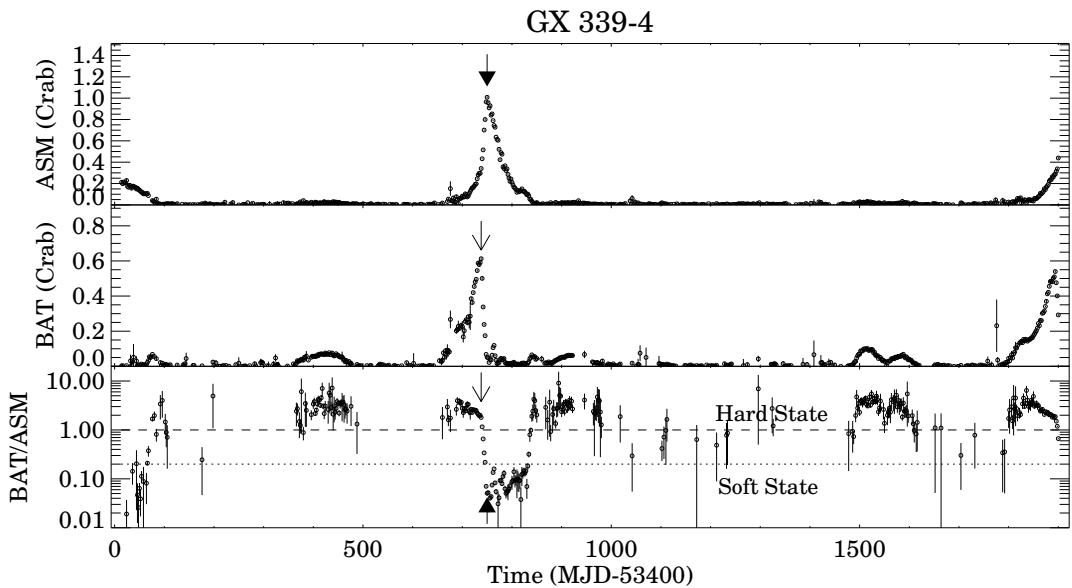}

   \begin{minipage}[]{150mm}

   \caption{ X-ray monitoring observations of black hole transient source GX 339-4 in 2-12 keV with the ASM and 15-50 keV with the BAT. A H-S transition was identified in the 2007 outburst. } \end{minipage}
   \label{Fig1}
   \end{figure}

   \begin{figure}[h!!!]
   \centering
   \includegraphics[width=8.0cm, angle=90]{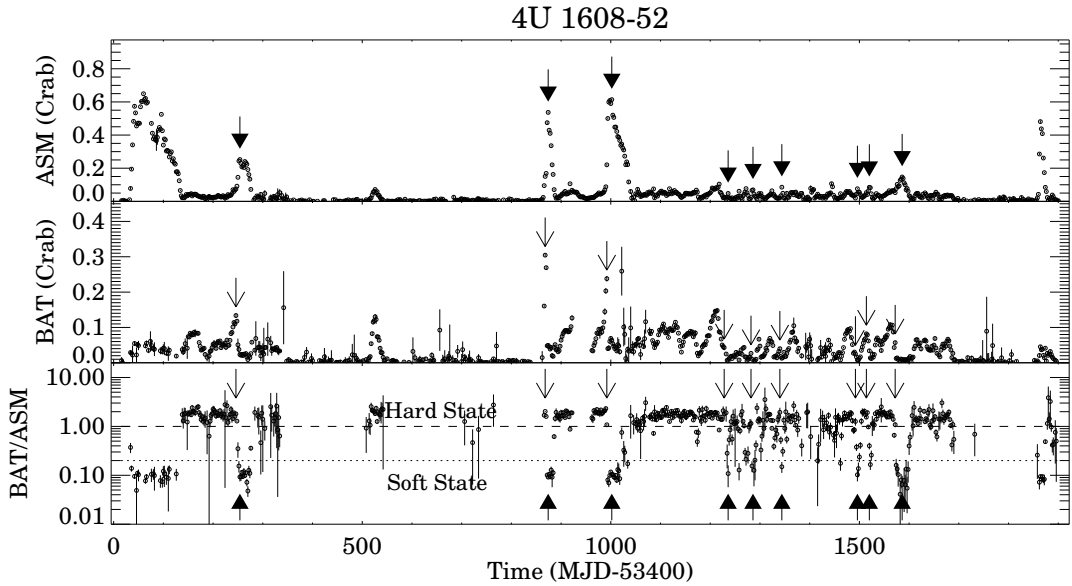}

   \begin{minipage}[]{150mm}

   \caption{ X-ray monitoring observations of neutron star transient source 4U 1608-52 in 2-12 keV with the ASM and 15-50 keV with the BAT. 6 more H-S transitions were identified. } \end{minipage}
   \label{Fig2}
   \end{figure}

   \begin{figure}[h!!!]
   \centering
   \includegraphics[width=8.0cm, angle=90]{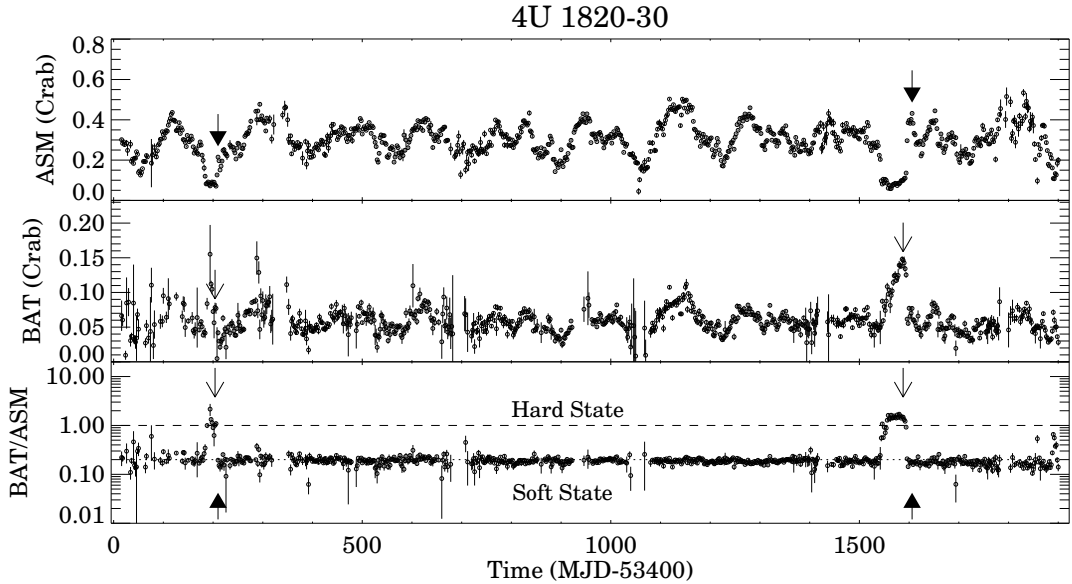}

   \begin{minipage}[]{150mm}

   \caption{ X-ray monitoring observations of neutron star persistent source 4U 1820-30 in 2-12 keV with the ASM and 15-50 keV with the BAT. A new H-S transition occurred in 2009 June. } \end{minipage}
   \label{Fig3}
   \end{figure}

\section{Automatic identification of state transitions}
\label{sect:mission}

We have completed a program, aiming at searching for state
transitions from the hard state to the soft state automatically.

At any time a source stayed in the hard state and at a later time the source is found in the soft state, there should have been a H-S transition during this period. These periods are identified as intervals of candidate transitions. Our program is aimed at the determination of the transition flux and the peak flux of the following soft state which is usually seen during the outburst/flare flux peak. We identify a state transition from the candidate transitions according to the following scenarios. Application of the following scenarios to the monitoring data before 2008 reproduces the results of Yu \& Yan (2009):

(1) The hardness ratio of the BAT flux peak we chose should correspond to
the hard state. That is, except that it was above 1.0, its
significance should be greater than $1\sigma$ above 0.6, the hardness ratio boundary between the hard state and the soft state (Yu \& Yan 2009). As to the peak flux of the following soft state, we chose the maximum ASM flux in the time range from the time corresponding to the first hardness ratio data below 1.0 to the time corresponding to the last hardness ratio data below 0.35 with the significance greater
than $1\sigma$ below 0.6. It was worth noting that in this time range, we
required there was at least one hardness ratio data below 0.2, and its
significance was greater than $1\sigma$ above 0 and below 0.35 respectively, to
make sure the source indeed transited to the soft state.

(2) We excluded those isolated ASM peaks as well as the BAT fluxes at the start of the transition detected only in a single time bin in order to avoid false identifications. We required that there was no data gap exceeding 4 days associated with the ASM peaks or the BAT peaks, since these data gaps prevent us from justifying the ASM peaks or the BAT peaks as flux peaks.

(3) Either the ASM peaks or the BAT peaks appearing as outliers defined below were
discarded. Outliers are defined as that the rate was 2 times larger than the rates of two adjacent data on either side for the ASM
data or 2.3 times for the BAT data. These values were chosen so that we can identify same transitions as found in Yu \& Yan (2009).

(4) When there was only one remarkable data point identified in the hard state
before a candidate transition, the hardness ratios of the adjacent points were required not
to be in the soft sate. Meanwhile, we required the significance of
the differences between this data and two adjacent data points in the hardness ratio
was smaller than $5\sigma$ to guarantee gradual spectral evolution on the time scale of 2-day bins.

(5) When there was only a single remarkable data point identified in the soft state,
the hardness ratios of the adjacent points were required not to be 3.5 times
larger than that of this point to guarantee gradual trend as (4).

Currently, our program can choose the peak flux of the soft state as the ASM
peak flux during the soft state period after the transition. In a few cases, the sources remain in the soft state and go up and down for a very long period, therefore the flux peak can not be physically associated with the state transition (as if the source would stay in the soft state forever, e.g., 4U 1820-30), whereas what we need
is the peak flux of the local flux peak immediately after the H-S transition.
This is done by a visual check. Besides, when the BAT peak flux appears during a H-S transition, we choose the maximal BAT flux in
the hard state range as the transition flux.

Our program can handle data from any other two instruments unless they cover the thermal and the non-thermal energy bands which can be used to identify state transitions.

\section{Results}
\label{sect:results}

In summary, we identified 128 H-S transitions in 20 neutron star LXMBs, 7 black hole LXMBs and 1 HMXB (Cyg X-3) between 2005 and 2010, while 59 H-S transitions are in the years 2008-2010. In these 28 sources, there are 7 new sources that had no transition in 2005-2008, including 3 black hole transients 4U 1630-47, SWIFT J1842.5-1124 and XTE J1752-223. Since the black hole masses and the distances of these 3 sources are unknown, they are not shown in the plot of the correlation between the luminosities in Eddington units.

In Figure~1--3, we show the long-term X-ray light curves of 3 bright Galactic X-ray binaries, in which spectral state transitions
were identified as examples. For each H-S transition, we marked the BAT flux
corresponding to the start of the transition with a thin arrow, and
the ASM peak flux of the following soft state with a thick arrow,
respectively. In these figures, we only plotted the hardness ratios with the significance greater than $1\sigma$ to make the figures clear.

In Figure~4, we plot the observed transition fluxes and the peak fluxes of the following soft state in 28 bright sources. It shows a strong positive
correlation, with a Spearman correlation coefficient of 0.85 and a chance
possibility on the order of $10^{-37}$. If we consider the data from
MJD 53413 to MJD 54504 used in Yu \& Yan (2009) and the data from
MJD 54504 to MJD 55304 respectively, we obtained the Spearman
coefficient of 0.85, with a chance possibility of $10^{-20}$, and the Spearman coefficient of 0.84, with a chance
possibility of $10^{-17}$, respectively.

To eliminate the effect due to diverse source distances and compact
star masses, we re-scaled the observed fluxes to intrinsic fluxes,
using the source distances and compact star masses with
uncertainties listed in Table~1. Meanwhile, assuming that the
Galactic binaries have similar spectral shape and Galactic hydrogen
absorption as the Crab, we convert the ASM flux and the BAT flux in
crab units into luminosities. Figure~5 shows the correlation between
the luminosity corresponding to the H-S transition and the peak
luminosity of the following soft state. The Spearman coefficient is
0.87, with a chance possibility of $1.2 \times 10^{-36}$ . We
fit the data with a model of the form $\log~L_{PS}=A~\log~L_{tr,H}+B$,
where $L_{PS}$ and $L_{tr,H}$ represent the outburst peak luminosity
in the soft state and the H-S transition luminosity (using the method in Kelly 2007). We obtained $A=0.96\pm0.05$ and
$B=0.52\pm0.08$, with an intrinsic scatter in $\log~L_{PS}$ of
$0.18\pm0.017$, taking into account the uncertainties in the
estimates of source distances and masses, if known. We excluded the
data of Cyg X-3 because it is uncertain whether it contains a black
hole or a neutron star. Likewise, to the former data, the Spearman
coefficient is 0.81, with a chance possibility of $1.9 \times
10^{-16}$. And $A=0.98\pm0.08$ and $B=0.55\pm0.12$, with an
intrinsic scatter in $\log~L_{PS}$ of $0.21\pm0.025$. While to the updated data, the Spearman coefficient is 0.90, with a chance
possibility of $2.2 \times 10^{-20}$. And $A=0.95\pm0.06$ and
$B=0.51\pm0.10$, with an intrinsic scatter in $\log~L_{PS}$ of
$0.16\pm0.025$.

Figure~6 shows the correlation between the luminosities in Eddington
units. The Spearman correlation coefficient is 0.87, with a chance
possibility of $2.2 \times 10^{-37}$ . We fit the data with the same
model above and obtained $A=0.88\pm0.06$ and  $B=0.36\pm0.11$, with
an intrinsic scatter in $\log~L_{PS}$ of $0.17\pm0.016$. Also, to
the former data, the Spearman coefficient is 0.83, with a chance
possibility of $1.6 \times 10^{-17}$. And $A=0.85\pm0.09$ and
$B=0.31\pm0.17$, with an intrinsic scatter in $\log~L_{PS}$ of
$0.20\pm0.024$. While to the updated data, the Spearman coefficient
is 0.90, with a chance possibility of $2.9 \times 10^{-20}$. And
$A=0.91\pm0.07$ and $B=0.41\pm0.15$, with an intrinsic scatter in
$\log~L_{PS}$ of $0.16\pm0.024$.

   \begin{figure}[h!!!]
   \centering
   \includegraphics[width=12.0cm, angle=0]{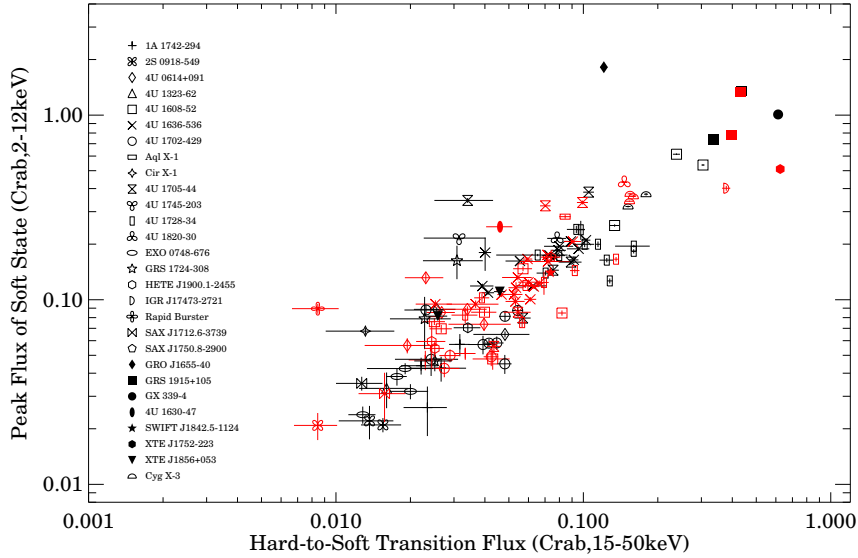}

   \begin{minipage}[]{130mm}

   \caption{ Observed BAT fluxes when the H-S transitions occured and the corresponding ASM peak fluxes of the following soft state. Hollow symbols and solid symbols represent neutron star and black hole binaries, respectively. Black symbols and red symbols represent data from MJD 53413 to MJD 54504 and data from MJD 54504 to MJD 55304, respectively. } \end{minipage}
   \label{Fig4}
   \end{figure}

   \begin{figure}[h!!!]
   \centering
   \includegraphics[width=12.0cm, angle=0]{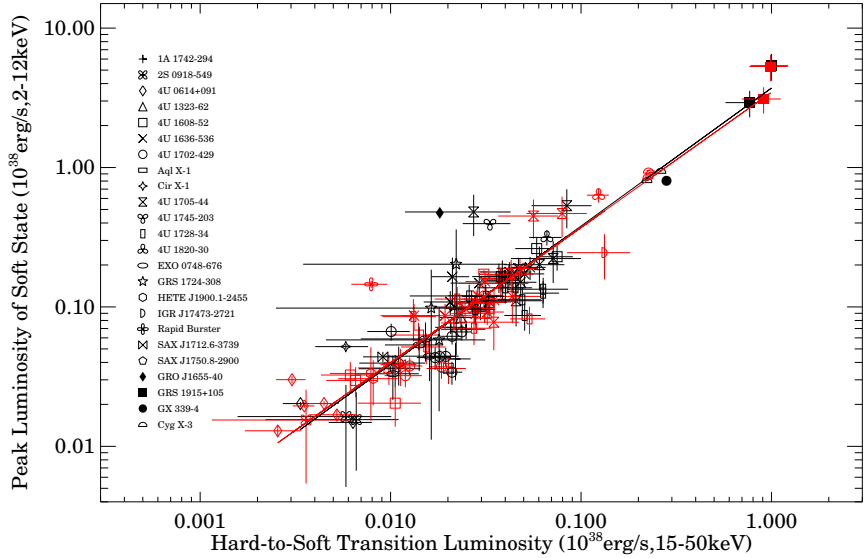}

   \begin{minipage}[]{130mm}

   \caption{ The transition luminosity(15-50keV,erg s$^{-1}$) and the peak luminosity of the following soft state(2-12keV,erg s$^{-1}$). Hollow symbols and solid symbols represent neutron star and black hole binaries, respectively. Black symbols and red symbols represent data from MJD 53413 to MJD 54504 and data from MJD 54504 to MJD 55304, respectively. The black line and the red line represent the fits of data from MJD 53413 to MJD 54504 and data from MJD 54504 to MJD 55304, respectively. } \end{minipage}
   \label{Fig5}
   \end{figure}

   \begin{figure}[h!!!]
   \centering
   \includegraphics[width=12.0cm, angle=0]{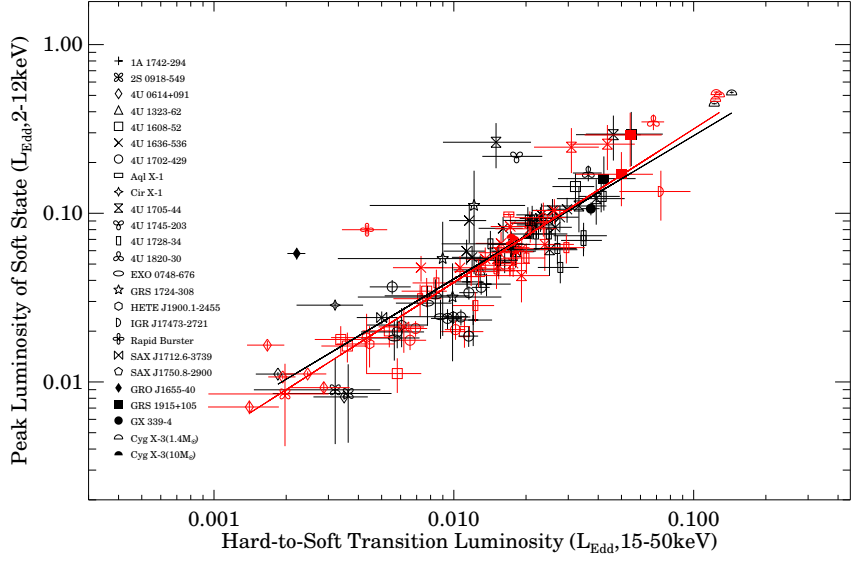}

   \begin{minipage}[]{130mm}

   \caption{  Correlation between the transition luminosity(15-50keV) and the peak luminosity of the following soft state(2-12keV) in Eddington units. Hollow symbols and solid symbols represent neutron star and black hole binaries, respectively. Black symbols and red symbols represent data from MJD 53413 to MJD 54504 and data from MJD 54504 to MJD 55304, respectively. The black line and the red line represent the fits of data from MJD 53413 to MJD 54504 and data from MJD 54504 to MJD 55304, respectively. } \end{minipage}
   \label{Fig6}
   \end{figure}

%
%               one-column-spanning table
%________________________________________ Table 2: Use_of_the routines

%% \input{table2.tex}
\begin{table}

\bc

\begin{minipage}[]{120mm}

\caption[]{ Statistics of the Sources with H--S Transitions Identified and
Parameters Used (Following Yu \& Yan
(2009))}\end{minipage}

\small
 \begin{tabular}{lcccclcc}
  \hline\noalign{\smallskip}
Source & Distance & Mass & H--S & new H--S & Ref.\\
   & (kpc) & ($\odot$) &  &  &  \\
  \hline\noalign{\smallskip}
1A 1742-294   & 8.0 & 1.4 & 6 & 2 & B\'{e}langer et al. (2006)\\
2S 0918-549   & 4.1--5.4 & 1.4 & 3 & 1 & in't Zand et al. (2005)\\
4U 0614+091   & 3 & 1.4 & 7 & 5 & Brandt et al. (1992)\\
4U 1323-62   & 10 & 1.4 & 1 & 0 & Parmar et al. (1989)\\
4U 1608-52   & $4.1\pm0.4$ & 1.4 & 9 & 6 & Galloway et al. (2008)\\
4U 1636-536   & $6\pm0.5$ & 1.4 & 28 & 15 & Galloway et al. (2008)\\
4U 1702-429   & $5.46\pm0.19$ & 1.4 & 10 & 3 & Galloway et al. (2008)\\
Aql X-1   & 5 & 1.4 & 1 & 1 & Rutledge et al. (2001)\\
Cir X-1   & 5.5 & 1.4 & 1 & 0 & Goss et al. (1977); Glass et al. (1994);\\
 & & & & & Jonker et al. (2004); Iaria et al. (2005)\\
4U 1705-44   & $7.4{+0.8 \atop -1.1}$ & 1.4 & 9 & 4 & Haberl~\&~Titarchuk (1995)\\
4U 1745-203   & 8.5 & 1.4 & 1 & 0 & Ortolani et al. (1994)\\
4U 1728-34   & $5.2\pm0.5$ & 1.4 & 18 & 7 & Galloway et al. (2008)\\
4U 1820-30   & $7.6\pm0.4$ & 1.4 & 2 & 1 & Kuulkers et al. (2003)\\
EXO 0748-676   & $7.4\pm0.9$ & 1.4 & 4 & 0 & Galloway et al. (2008); Wolff et al. (2005)\\
GRS 1724-308   & $7\pm2$ & 1.4 & 3 & 0 & Galloway et al. (2008)\\
HETE J1900.1-2455   & $4.7\pm0.6$ & 1.4 & 3 & 2 & Galloway et al. (2008)\\
IGR J17473-2721 & $4.9\pm0.8$ & 1.4 & 1 & 1 & Altamiraino et al. (2008) \\
Rapid Burster   & 8 & 1.4 & 1 & 1 & Ortolani et al. (1996)\\
SAX J1712.6-3739   & 7 & 1.4 & 2 & 1 & Cocchi et al. (2001)\\
%SAX J1747.0-2853   & $7.5\pm1.3$ & 1.4 & 1 & 0 & Werner et al. (2004)\\
SAX J1750.8-2900   & $6.3\pm0.7$ & 1.4 & 1 & 1 & Kaaret et al. (2002); Natalucci et al. (1999)\\
GRO J1655-40   & 3.2 & $6.3\pm0.5$ & 1 & 0 & Greene et al. (2001); Hjellming~\&~Rupen (1995)\\
GRS 1915+105   & 11.2-12.5 & $14\pm4$ & 4 & 2 & Greiner et al. (2001); Fender et al. (1999);\\
 & & & & & Mirable~\&~Rodriguez (1994)\\
GX 339-4   & $\ge5.6$ & $\ge5.8\pm0.5$ & 1 & 0 & Hynes et al. (2003); Shahbaz et al. (2001)\\
4U 1630-47   & Unknown & Unknown & 1 & 1 & \\
SWIFT J1842.5-1124   & Unknown & Unknown & 1 & 1 & \\
XTE J1752-223   & Unknown & Unknown & 1 & 1 & \\
XTE J1856+053   & Unknown & Unknown & 2 & 0 & \\
Cyg X-3   & 10 & Unknown & 5 & 3 & Dickey (1983)\\
  \noalign{\smallskip}\hline
\end{tabular}
\ec
%% place \tablecomments and \tablerefs below \end{center| and \end{center}:
%% you may leave the table-width parameter to editors or set to your actual size.
\tablecomments{0.86\textwidth}{H-S: H-S transitions discovered from
MJD 53413 to MJD 55304. new H-S: H-S transitions discovered from MJD
54504 to MJD 55304. For neutron stars, no accurate mass measurement
is known. So 1.4 solar masses were used. For GX 339-4, only lower
limits of the distance and the black hole mass, 5.6 kpc and 5.8
solar masses, are known. These values were used as the actual
distance and mass.}
\end{table}

\section{Conclusions and discussion}
\label{sect:conclusion}

Using X-ray monitoring observations with the
RXTE/ASM and the Swift/BAT, we have performed an automatic search for the H-S state transitions in
bright persistent and transient X-ray binaries during a
period of five years from 2005 to 2010. The identification of transitions and the correlation between the transition luminosity and the outburst/flare peak luminosity in the following soft state obtained with our automatic routine from the data in 2005-2008 is consistent with that reported in Yu \& Yan (2009). From the recent observations in 2008-2010, we have identified 59 more H-S state transitions in 28 Galactic X-ray binaries, which also show the same positive correlation.

\subsection{H-S Transitions Newly Identified in Individual Sources}

Let's take a closer look at the new sources other than those in Yu \& Yan (2009). In Aql X-1, at least 4 H-S transitions were seen before 2005 (Yu \& Dolence 2007), but none in 2005-2008, then a new one occurred at the end of 2009. IGR J17473-2721 was only observed a week outburst in 2005, characterized by a hard state spectrum. A H-S state transition was seen in 2008 June. 4U 1630-47 is one of the most active black hole transients, and it has produced strong hard X-ray emission during its 17 detected outbursts (Tomsick et al. 2005). The latest outburst started in 2009 December, underwent the H-S transition, and went back to the hard state until 2010 August (Tomsick \& Yamaoka 2010). SWIFT J1842.5-1124 was first detected in 2008 July (Krimm et al. 2008). Its X-ray spectral and timing properties are similar to the black hole in the hard state (Markwardt et al. 2008). A H-S transition occurred in 2008 September according to the BAT and ASM light curves. XTE J1752-223 was discovered during RXTE scanning the Galactic bulge region on 23 October 2009 (Markwardt et al. 2009) and there is strong evidence indicating that it is a black hole X-ray transient. Since the discovery of the source, it showed little spectral evolution and had been in the hard state. Until 2010 January, it underwent a H-S transition (Homan 2010). In addition, 4U 1820-30 remained in the soft state after the recent H-S transition in 2005 August. In 2009 April, it entered the hard state and after two months, another H-S transition took place. This phenomenon is atypical for 4U 1820-30 over the past 10-15 years, since the source usually stays in the hard state for one to two weeks (Krimm et al. 2009).

Furthermore, by analyzing H-S transitions during 2005-2010 in single sources, we have found that the transition luminosities of 4U 1608-52 in recent two years were much lower than that in 2005-2008, different from the brightest H-S transition by an order of magnitude. Similar large variation of the transition luminosity has also been seen previously in Aql X-1 (Yu \& Dolence 2007). Besides, in the single source 4U 1636-53, which had experienced many spectral transitions during the studied time period, there exists the correlation between the luminosities in Eddington units well. Such a correlation has been found in individual sources XTE J1550-564, Aql X-1 and 4U 1705-44 (Yu et al. 2004). Detailed study of the 4U 1608-52 and 4U 1636-53 will be presented elsewhere.

\subsection{Estimates of the Significance of the H-S Transitions Identified}

Yu \& Yan (2009) studied the duration of the bright Galactic X-ray binaries staying at certain hardness ratios and found the hardness ratios of the hard state and the soft state are in the range above 0.6 and below 0.35, respectively, showing distinct bimodal distribution. Therefore we used the two thresholds to estimate the significance of the H-S transitions we identified. We used the data point immediately before the transition whose hardness ratio is above 1.0 and the preceding two data points to calculate the significance that at least one 2-day hardness ratio is greater than 0.6 --- the significance that the source was in the hard state. Similarly, we used the first data point following whose hardness ratio is below 0.2 and the following two data points to calculate the significance that at least one 2-day hardness ratio is smaller than 0.35 --- the significance that it is in the soft state. A H-S transition should have occurred in this interval if the soft state appears after the hard state. The result was that in these 128 transitions, 125 exceeds $2\sigma$ and 101 exceeds $3\sigma$. If we added one data point following the hard state data and one data point preceding the soft state data to calculate the significance of reaching the hard state and the soft state respectively, there are 126 out of these 128 transitions larger than $2\sigma$ and 105 larger than $3\sigma$. Thus, most of the identifications of state transitions we identified are reliable.

\subsection{Future Work}

After including 2008--2010 data compared with the data used in Yu \& Yan (2009), the H-S transition luminosities remain to spread over a luminosity range of about 2 orders of magnitude, showing no cutoff or saturation at either end of the correlation. The large span of the transition luminosities indicates that using mass accretion rate to interpret the state transition is not correct. According to the conclusion of Yu \& Yan (2009), the rate of increase of the mass accretion rate drives the H-S transition in most of the state transitions in X-ray binaries. 2S 0918-549 and 4U 0614+091, which have the lowest transition
luminosities, can be explained as that their $\frac{d\dot{M}}{dt}$ is the least. And Cyg X-3, if it contains a neutron star, has the highest transition luminosities because of the highest $\frac{d\dot{M}}{dt}$ (see
Figure 26. in Yu \& Yan 2009). In general, transient sources tend
to have higher $\frac{d\dot{M}}{dt}$ than persistent sources and
therefore higher H-S transition luminosities.

%The correlation has been found in individual sources XTE J1550-564, Aql X-1 and 4U 1705-44 (Yu et al. 2004). Taking a further analysis of single sources with these data, we have found that the transition luminosities of 4U 1608-52 in recent two years were much lower than that in 2005-2008, different from the brightest H-S transition by an order of magnitude. Similar large variation of the transition luminosity has also been seen previously in Aql X-1 (Yu \& Dolence 2007). Besides, 4U 1636-53, which had experienced many spectral transitions, shows the correlation between the luminosities in Eddington units well. Detailed study of the two sources individually with the help of pointed observations will be presented elsewhere.

Modification of the program will allow us to automatically identify H-S transitions using monitoring data other than the BAT and the ASM, for example, MAXI, unless the instruments cover the soft component and the non-thermal components. Noting that the hardness
ratio thresholds for the hard state and the soft state should be
determined first, just as Yu \& Yan (2009), for the hardness ratio range of the hard state and the soft state will change when we use different instruments or different energy bands to calculate the hardness ratio. Additionally, the program can be modified to search for the soft-to-hard transition in the outburst or flare in which the H-S transition was identified. Application to instant alert or predict state transitions and to study the hysteresis effect of spectral state transitions is under further investigation.

\normalem
\begin{acknowledgements}
We thank the RXTE and the Swift Guest Observer Facilities at NASA
Goddard Space Flight Center for providing the RXTE/ASM products and
the Swift/BAT transient monitoring results. We would like to thank the anonymous referee for useful comments. WY appreciate useful discussions with Mike Nowak, Chris Done, Feng Yuan, Robert Fender, Nan Shuang Zhang and Robert Soria. This work was supported
in part by the National Natural Science Foundation of China (10773023, 10833002,11073043), the One Hundred Talents project of the Chinese Academy of Sciences, the Shanghai Pujiang Program (08PJ14111), the National Basic Research Program of China (973 project 2009CB824800), the grant for concept study of space science from the Chinese Academy of Science, and the starting funds at the Shanghai Astronomical Observatory. The study has made use of data obtained through the High Energy Astrophysics Science Archive Research Center Online Service, provided by the NASA/Goddard Space Flight Center.
\end{acknowledgements}

\label{lastpage}

\end{document}